\documentclass[a4paper, oneside, 12pt]{scrartcl}

\usepackage[utf8x]{inputenc}
\usepackage[T1]{fontenc}

\usepackage[english]{babel}

\makeatletter
\DeclareOldFontCommand{\rm}{\normalfont\rmfamily}{\mathrm}
\DeclareOldFontCommand{\sf}{\normalfont\sffamily}{\mathsf}
\DeclareOldFontCommand{\tt}{\normalfont\ttfamily}{\mathtt}
\DeclareOldFontCommand{\bf}{\normalfont\bfseries}{\mathbf}
\DeclareOldFontCommand{\it}{\normalfont\itshape}{\mathit}
\DeclareOldFontCommand{\sl}{\normalfont\slshape}{\@nomath\sl}
\DeclareOldFontCommand{\sc}{\normalfont\scshape}{\@nomath\sc}
\makeatother

\usepackage[a4paper,  centering, bindingoffset=0mm, inner=25mm, outer=25mm, top=26mm, bottom=36mm, heightrounded]{geometry}

\usepackage{graphicx}
\usepackage{float}
\usepackage{cite}

\usepackage{amsmath}	
\usepackage{amssymb}
\usepackage{amsfonts}
\usepackage{mathrsfs}
\usepackage{mathtools}
\usepackage{amsthm}
\usepackage{esint}     

\usepackage{braket}
\usepackage{slashed}

\usepackage[font=small,labelfont=bf, width=.95\textwidth]{caption}

\usepackage{verbatim}
\usepackage{textcomp}
\usepackage{enumitem}

\usepackage[toc]{appendix}

\usepackage{layout}
\usepackage{microtype}
\usepackage{bookmark}

\usepackage{color}
\usepackage[]{xcolor}

\newcommand{\dd}{d} 
\newcommand{\ii}{\mathrm{i}}

\makeatletter
\newcommand{\subalign}[1]{%
  \vcenter{%
    \Let@ \restore@math@cr \default@tag
    \baselineskip\fontdimen10 \scriptfont\tw@
    \advance\baselineskip\fontdimen12 \scriptfont\tw@
    \lineskip\thr@@\fontdimen8 \scriptfont\thr@@
    \lineskiplimit\lineskip
    \ialign{\hfil$\m@th\scriptstyle##$&$\m@th\scriptstyle{}##$\crcr
      #1\crcr
    }%
  }
}
\makeatother

\definecolor{unamblue}{cmyk}{1 0.79 0.12 0.59}

\usepackage[]{hyperref}
\hypersetup{
    colorlinks=true,%
    citecolor=unamblue,%
    filecolor=unamblue,%
    linkcolor=unamblue,%
    urlcolor=unamblue,
    bookmarksnumbered=true,     
    bookmarksopen=true,         
    bookmarksopenlevel=1,       
    pdfstartview=Fit,           
    pdfpagemode=UseOutlines,
    pdfpagelayout=TwoPageRight
}

\usepackage{graphicx}
\usepackage{tikz}
\usepackage{slashed}
\usepackage{epstopdf}
\usepackage{verbatim}	
\usepackage{blkarray}
\usepackage{tikz}
\usepackage{ytableau}
\usepackage{tikz-feynman}
\tikzfeynmanset{warn luatex=false}

\newcommand{\cc}{c}

\usepackage{subcaption}

\newcommand{\be}{\begin{equation}}
\newcommand{\ee}{\end{equation}}

\newcommand{\la}{\langle}
\newcommand{\ra}{\rangle}

\newcommand{\ba}{\begin{equation} \begin{aligned}}
\newcommand{\ea}{\end{aligned} \end{equation}}

\def\del{\Delta}

\title{\Huge  Worldline description of a bi-adjoint scalar
and the zeroth copy
\\}

\author{ \normalfont\normalsize Fiorenzo Bastianelli, Francesco Comberiati, Leonardo de la Cruz\\[2mm]
\emph{\normalfont\small \em Dipartamento di Fisica e Astronomia ``Augusto Righi'', Universit\`a di Bologna}\\
\emph{\normalfont\small \em and INFN Sezione di Bologna, via Irnerio 46, I-40126 Bologna, Italy}	
 }

\date{%
 $\,$%
    \\[2\baselineskip]
    \normalfont\normalsize%
      \parbox{0.8\linewidth}{%
{\bf \sf Abstract}. 
Bi-adjoint scalars are helpful in studying properties of color/kinematics duality and the double copy, which relates scattering amplitudes
of gauge and gravity theories. Here we study bi-adjoint scalars from a worldline perspective. 
We show how a global $G\times \tilde G$ symmetry group may be realized by worldline degrees of freedom. 
The worldline action gives rise to vertex operators, which are compared to similar ones 
describing the coupling to  gauge fields and gravity, thus exposing the color/kinematics interplay in this framework.
The action is quantized by path integrals to find a worldline representation of the 
one-loop QFT effective action of the bi-adjoint scalar cubic theory.
As simple applications, we recover the one-loop beta function 
of the theory in six dimensions, verifying its vanishing, and compute the self-energy correction to the propagator. 
The model is easily extendable to that of a particle carrying  
an arbitrary representation of direct products of global symmetry groups, including the 
multi-adjoint particle, whose one-loop beta function we reproduce as well.}
}

\begin{document}

\maketitle
\thispagestyle{empty}

\newpage

\section{Introduction}
The duality between color and kinematics \cite{Bern_2008,Bern:2010ue} reveals a deep structure relating amplitudes 
of gauge and gravity theories,  see \cite{Bern:2019prr} for a review.
An interesting model, used for studying color/kinematic relations, is that of a bi-adjoint scalar 
\cite{Cachazo:2013hca, Cachazo:2013iea}.  Its basic structure is that of a $\phi^3$ theory, a standard QFT toy model which is 
super-renormalizable in 4 dimensions and renormalizable in 6 dimensions (where it becomes asymptotically free),
but with the scalar field extended to be symmetric under the action of a compact group of the form $G\times \tilde G$. In particular,  
the scalar is taken to transform in the adjoint representation for each factor, 
so that its Feynman rules carry 
corresponding symmetry factors (named color factors). The latter may be substituted by kinematical factors having the same symmetries,  thus producing  Feynman rules and amplitudes of non-abelian gauge fields and gravity.

In a purely classical setting, the bi-adjoint scalar has  also  served as the basis to find maps between classical solutions  in gauge and gravity at both non-perturbative  \cite{Monteiro:2014cda, Luna:2015paa,Luna:2016due, White:2016jzc}  and  perturbative  \cite{Goldberger:2016iau, Goldberger:2017frp, Goldberger:2017vcg, Shen:2018ebu} levels. The perturbative solutions of the bi-adjoint scalar theory coupled with point
sources (parametrized by their worldline coordinates) are obtained by iteratively solving the equations of motion. 
The kinematic structure of these solutions is crucial for a systematic amplitude-like implementation of the double copy \cite{Shen:2018ebu}.  
At classical perturbative level, the form of  the action leading to equations of motion is unimportant as long as it reproduces the equations of motion, but its precise form is central to undertake quantization.

Seeds of color/kinematic relations, and corresponding double copy structure that reproduces 
gravity amplitudes from the gauge ones, can be traced back to the origin of string theory,
when it was noticed that the Veneziano amplitude  (an open-string scattering amplitude) could be related 
in a simple way to the Virasoro-Shapiro amplitude (a closed-string amplitude).
Open strings give rise to gauge fields while closed strings give rise to gravitons, 
so one notices a signal of the gauge/gravity relations \cite{Kawai:1985xq}. The string results were obtained and interpreted  in a first quantized picture.   Similarly, in this paper we wish to investigate the color/kinematic relations using a first quantized approach to particle theory \cite{Schubert:2001he}.

Thus, after some preliminary
considerations, we construct a worldline action  for the  bi-adjoint particle, paying attention to the realization of its color degrees of freedom.
The traditional approach is to use a matrix-valued particle action, whose matrix indices act directly on the indices 
of the wave-function, defined ab-initio to have multiple components which correspond to the color degrees of freedom 
of the particle\footnote{Strictly speaking, as long as the color is not gauged, one should name them flavor degrees of freedom, 
but we keep with the current terminology.}.
This approach requires a path ordering prescription for the path integral quantization, as 
 in \cite{Bastianelli:1992ct} where worldline methods were employed to compute trace anomalies.

An alternative approach is to introduce color variables, using
the set-up described in  \cite{Bastianelli:2013pta, Bastianelli:2015iba, Ahmadiniaz:2015xoa},
where  color variables are coupled to a worldline  $U(1)$  gauge field
which carries an additional Chern-Simons term. 
The gauge field implements a projection to the desired representation of the symmetry group
assigned to the particle. The choice of the representation is obtained by tuning the Chern-Simons coupling to a suitable discrete value. A form of this projection was described also in \cite{DHoker:1995uyv}, and later recognized to arise from the coupling to  a $U(1)$ gauge field. 
The projection mechanism was seen at work in the worldline 
path integral treatments  of differential forms \cite{Bastianelli:2005vk, Bastianelli:2011pe, Bastianelli:2012nh}, 
and extended shortly afterwards 
to color degrees  of freedom  \cite{Bastianelli:2013pta}.  
Further extensions have been discussed in 
\cite{Corradini:2016czo, Edwards:2016acz}.
On the other hand, worldline color variables had been introduced  
originally much earlier \cite{Balachandran:1976ya, Barducci:1976xq}. 

While the two methods are equivalent, the
color variable approach may turn out to be 
more useful for studying the classical limit of the color charge, as its leaves 
open the option of projecting to different sectors of the Hilbert space, and in particular to those  containing 
a large
number of copies of the original elementary charge. We will  not address this issue here, though.

 After having identified the correct worldline action for the bi-adjoint particle,
we use it to represent the one-loop effective action of the bi-adjoint scalar 
 field. Then, we 
extract the one-loop beta function of the theory in 6 dimensions, which was
recently shown to vanish \cite{Gracey:2020baa}. 
We also study  the one-loop self-energy correction
to the bi-adjoint  propagator.  Eventually, we indicate how our model can be extended  
to that of a particle carrying  an arbitrary representation of direct products of global symmetry groups, and compute 
the one-loop beta function of a multi-adjoint scalar particle in six dimensions.  

The rest of the paper  is organized as follows: In Section \ref{sec2}, after reviewing our motivations,
 we derive two alternative forms of the action that describes a bi-adjoint scalar particle. 
In Section \ref{effective-action} we discuss the corresponding one-loop effective action
using these two alternative  but equivalent 
approaches, and give examples. Section \ref{extensions} discusses extensions of the 
bi-adjoint model to more general situations. Our Conclusions are given in Section \ref{Conclusions}.  
  
\section{Construction of worldline actions}
\label{sec2}
\subsection{Motivation: from gauge to scalar and gravity }
A recurrent topic in the worldline approach to quantum field theory is the derivation of parameter integrals which sum up whole classes of
Feynman diagrams. At 1-loop one is sometimes interested in obtaining a 1PI effective action induced by a scalar loop in 
scalar/gauge/graviton field backgrounds, $\Psi^{(s)}(x)$, where $s=0,1,2$.  Denoting by $\chi$ a set of auxiliary variables 
for the particle, the effective action has the schematic  form
\begin{align}
 \Gamma[\Psi^{(s)}]
 = \int_{S^1} \frac{Dx D \chi }{\text{Vol} (\text{Gauge})} e^{-S[x,\chi;\Psi^{(s)}(x)]} \;,
 \label{schematic-action}
\end{align}
where $S[x,\chi;\Psi^{(s)}(x)]$ is the Euclidean action for a scalar particle coupled to a background field  $\Psi^{(s)}$, 
and $S^1$ is the circle (the loop). The auxiliary variables $\chi$ may include the einbein $e$, color variables, or ghosts. 
After setting appropriate boundary conditions and performing a gauge-fixing, the path integral can
be solved perturbatively by considering Wick contractions of vertex operators as follows
\begin{align}
 \Big\langle V^{(s)}[k_1, \varepsilon_1; x_1,\chi_1] \cdots V^{(s)} [k_n, \varepsilon_n; x_n,\chi_n] \Big\rangle \;,
\end{align}
where the total number $n$ corresponds to the total number of plane waves of the external background fields 
($k_i$ and $\varepsilon_i$ denote their momenta and polarizations, $x_i \equiv x(\tau_i)$, etc.).

For instance, the gluon vertex operator corresponding to $s=1$ is given by \cite{Bastianelli:2013pta}
\begin{align}
 V^{(1)}[k,\varepsilon, a; x, \cc, \bar \cc]= \int_0^1 \dd \tau  \ \varepsilon \cdot  \dot x (\tau)  \
\bar \cc_{a'}(\tau)  (T^a)^{a'}{}_{b'}\, \cc^{b'}(\tau) \  e^{\ii k\cdot x(\tau)} \;,
\label{gluon-vertex}
\end{align}
where $\cc^{a'}(\tau)$ and $\bar \cc_{a'}(\tau)$  are auxiliary worldline color variables 
transforming in the fundamental and antifundamental representations of the gauge group $G$, and 
$(T^a)^{a'}{}_{b'}$ are the generators in the fundamental representation.
Other representations could be considered as well, of course. In the spirit of the double copy, we may define the kinematic factor 
  $ K(\tau)=\varepsilon \cdot  \dot x (\tau)$ and the color factor
    $C^a(\tau)=\bar \cc_{a'}(\tau) (T^a)^{a'}{}_{b'}\, \cc^{b'}(\tau)  $, and 
consider in \eqref{gluon-vertex}  the replacement 
\begin{align}
 C^a(\tau) \ \rightarrow \ \tilde{K}(\tau) \;,
 \label{prescription-1}
\end{align}
where the tilde indicates dependence on another polarization vector $\tilde \varepsilon$. 
We obtain the following vertex operator 
\begin{align}
 V^{(2)}[k,\varepsilon; x]= \int_0^1 \dd \tau  \ \varepsilon \cdot  \dot x (\tau)  \
\tilde \varepsilon \cdot  \dot x (\tau) \  e^{\ii k\cdot x(\tau)} \;,
\label{scalar-grav-vertex}
\end{align}
which should correspond to a vertex operator for the emission/absorption of a graviton 
\cite{Bastianelli:2002fv}.
This is indeed correct, after identifying the graviton polarization by
$\varepsilon_\mu \tilde\varepsilon_\nu \rightarrow \varepsilon_{\mu\nu}$, and eventually taking into account 
prescriptions for regulating UV ambiguities on the worldline (this is typical for the coupling of a particle 
to gravity \cite{Bastianelli:2006rx, Bastianelli:2011cc}).

Now, inspired by the zeroth copy, let us consider instead the replacement
\begin{align}
 K(\tau) \ \rightarrow \ \tilde{C}^{\alpha}(\tau) \;,
\label{prescription-2}
\end{align}
with a color factor associated to a different symmetry group $\tilde G$,
carrying its own color variables $d^{\alpha'}(\tau)$ and  $\bar d_{\alpha'}(\tau)$ taken in the fundamental and antifundamental 
representation, respectively. This replacement produces the vertex operator
\begin{align}
  V^{(0)}[k, a, \alpha;  x, \cc, \bar \cc, d, \bar d]= \int_0^1 \dd \tau  \
  \bar \cc_{a'}(\tau)  (T^a)^{a'}{}_{b'}\, \cc^{b'}(\tau) \
  \bar d_{\alpha'}(\tau)  (T^\alpha)^{\alpha'}{}_{\beta'}\, d^{\beta'}(\tau) \
     e^{\ii k\cdot x(\tau)} \;,
  \label{bi-adjoint-vertex}
\end{align}
where $a$ and $\alpha$ are indices of the adjoint representation of $G$ and $\tilde G$, respectively. In the 
case of scattering amplitudes it is well-known that the replacement of kinematic numerators by 
color ones leads to the bi-adjoint scalar theory \cite{Weinzierl:2016bus, Cheung:2017pzi}. Here, we see that this replacement produces a vertex 
operator for the  coupling  of the particle to the plane wave of a bi-adjoint scalar, 
reminding the classical double copy of \cite{Monteiro:2014cda}. This vertex operator can be used to  
obtain an effective action of the form \eqref{schematic-action}
for a particle coupled to a background bi-adjoint scalar field $\Psi^{(0)} \sim \Phi^{a \alpha}$.

\subsection{Worldline  matrix-valued action}
\label{Wmva}

The previous considerations motivate us to derive in more details the worldline action for a particle coupling to the 
bi-adjoint scalar $\Phi^{a\alpha}$. We consider directly  the special case of  
a particle interpreted as the quantum of the bi-adjoint scalar field itself. For that purpose, we start from 
the QFT of a bi-adjoint scalar field $\Phi^{a \alpha}$, charged under a $G\times \tilde G$ global symmetry group 
and transforming in the adjoint for each factor\footnote{The Lie algebra for each factor has the form 
$ [T^a,T^b]=i f^{ab}{}_c  T^c $ and the adjoint representation is given by the structure constants
$(T_{\rm A}^a)^b{}_c= -if^{ab}{}_c = -if^{abc}  $. The Killing metric is normalized to  
$\delta^{ab}$, so that upper and lower adjoint indices are equivalent. We use greek indices to label generators of the  
second group $\tilde G$.}.
The Euclidean Lagrangian of the model is  \cite{Luna:2015paa,White:2016jzc}
  \be
{\cal L } = \frac12 \partial_\mu \Phi^{a \alpha} \partial^\mu \Phi^{a \alpha} 
+\frac{m^2}{2} \Phi^{a \alpha}\Phi^{a \alpha} 
+\frac{y}{3!} f^{abc} \tilde f^{\alpha \beta \gamma}
\Phi^{a \alpha} \Phi^{b \beta} \Phi^{c \gamma} \:, 
\label{lag-phi3}
\ee
where $m$ is the mass and $y$ the coupling constant. This model is the massive version of a massless one, whose amplitudes in the
Cachazo-He-Yuan (CHY) representation were considered in \cite{Cachazo:2013hca, Cachazo:2013iea}. The 
massive generalization of these amplitudes was considered e.g., in \cite{Naculich:2014naa, delaCruz:2016wbr, Brown:2018wss}.  In a quantum/background split ($\Phi^{a \alpha} \to \phi^{a \alpha} + \Phi^{a \alpha}$) 
one finds, after a partial integration, a quadratic action for the quantum field  $\phi^{a \alpha}$ 
 \be
{\cal L }^{(2)}_q = \frac12 \phi^{a \alpha} \Big( \delta^{ab}\delta^{\alpha \beta} (-\partial^2 + m^2)
+ y f^{abc}\tilde f^{\alpha\beta \gamma} \Phi^{c \gamma}  \Big) \phi^{b \beta} \;, 
 \label{quad-act} 
 \ee
which is  coupled to the background $\Phi^{a \alpha}$. It contains a quadratic differential operator  of the form
(setting $p^2= -  \partial^2 $)
\be
2 H =
\delta^{ab}\delta^{\alpha\beta} (p^2+m^2)  + y f^{abc} \tilde f^{\alpha\beta\gamma} \Phi^{c \gamma} \;,
\label{ham-1} 
\ee
which is interpreted as a particle Hamiltonian $H$--- the factor 2 being conventional --- which 
acts on wave functions $\phi^{a \alpha}(x)$ that carry indices of the bi-adjoint. The inverse of this Hamiltonian 
gives the bi-adjoint propagator in the  $\Phi^{a \alpha}$ background,
and the functional trace of the corresponding heat kernel is related to the one-loop effective action \cite{Schwinger:1951nm}.
As it stands, this Hamiltonian has a matrix-valued potential, and it produces a matrix-valued action. 
In a path integral, a path ordering must be used to properly define its exponential.

To present the particle Hamiltonian in a more compact way,  let us 
introduce a matrix notation to cast \eqref{quad-act}  in the form
 \be
{\cal L}_q^{(2)} = \frac 12 \phi^T \left (-\partial^2 + m^2 + y \hat \Phi  \right) \phi \;,
\label{43}
\ee 
where $\phi = \phi^A= \phi^{a \alpha}$ has to be considered as a column vector with components labeled by the multi-index
$A= (a\alpha)$, while  
$\hat \Phi$ is a matrix-valued background potential 
\be
\hat \Phi = \hat \Phi^{A, B}=  
\hat \Phi^{a\alpha, b \beta}= 
f^{abc} \tilde f^{\alpha \beta \gamma} \Phi^{c \gamma} \;.
\ee
The matrix $\hat \Phi$ is symmetric in the indices $A$ and $B$.
The corresponding matrix-valued Hamiltonian, eq.~\eqref{ham-1}, is then cast as 
\be
2 H = p^2 + m^2 + y \hat \Phi(x) \;. 
\ee
It acts on wave functions of the form $\phi(x) = \phi^A(x)$, which have components labeled by the multi-index $A$.
The Hamiltonian leads classically to an action that is also matrix-valued.
The latter, after some simple redefinitions, can be written in the form 
\be
 S[x]=
\int_0^1 \dd\tau 
\Big [ \frac{1}{4 T} \dot x^2 + Tm^2 +T  y \hat \Phi(x)   \Big ] ,
\label{2.14} 
\ee
where $T$ is the Fock-Schwinger proper-time. The corresponding path integral needs a time ordering prescription, that makes sure that it
solves the correct Schr\"odinger equation (just as for time dependent Hamiltonians).
Then,  following Schwinger \cite{Schwinger:1951nm},
 we find that the worldline path integral for the
effective action \eqref{pi-2} takes the form

\begin{align}
\Gamma[\Phi] = -\frac12 
\int_0^\infty \frac{\dd T}{T} e^{-m^2T}\,
{\rm tr} \int_{\cal P} Dx \, 
{\bf T} \, e^{-S[x]} \;,
\label{pi-new1}
\end{align}
where $\bf T$ denotes the time ordering, 
`${\rm tr}$' the additional trace on the finite color degrees of freedom, and with the constant mass term
explicitly taken out of the worldline action $S$.  The symbol $\cal P$ indicates that the coordinates $x$ must have periodic boundary conditions in order to produce the trace in the corresponding Hilbert space.  
\subsection{Worldline action  with auxiliary color variables}

A way to get rid of path ordering it to use color variables. The quantization of the latter gives rise to 
creation/annihilation operators that create color degrees of freedom in the Hilbert space of the particle.
At the same time their worldline propagator reconstructs the path ordering.
Eventually, a projection on a fixed occupation number of the color degrees of freedom
selects the precise representation of the symmetry group (the color charge)
that one is willing to assign to the particle. This last step is achieved  by coupling the color variables
to a $U(1)$ worldline gauge field with an additional discrete Chern-Simons coupling, 
that fixes  the chosen color occupation number in the Hilbert space \cite{Bastianelli:2013pta}.
Thus, recalling that for a relativistic particle the Hamiltonian must be gauged\footnote{ 
The corresponding gauge field (the einbein) will eventually lead to the Fock-Schwinger proper time
of the previous  section.}, 
one is led to the following worldline action in phase space (subscript `$\text{ps}$') for the bi-adjoint scalar in real time $\tau$
\be
S_{\text{ps}} =\int \dd\tau 
\Big [ p_\mu \dot x^\mu + \ii \bar \cc_a \dot \cc^a  + \ii \bar d_\alpha \dot d^\alpha  
-e H - a J -  {\tilde a}{\tilde  J}  \Big ]  \;,
\label{action1}
\ee
where $H, J,\tilde J$ denote first class constraints 
\begin{align}
        H &  = \frac12 ( p^2+ m^2 -y\, C^a  \Phi^{a\alpha}(x) {\tilde C}^{\alpha} ) \;, \qquad 
       C^a= -\ii f^{abc} \bar c^b c^c \;,
        \quad {\tilde C}^\alpha = -\ii \tilde f^{\alpha\beta\gamma} \bar d^\beta d^\gamma \;,
 \label{ham}\\
        J &  =  \bar \cc_a \cc^a -s \;, 
       \label{cons1}  
        \\
            \tilde J   &= \bar d_{\alpha} d^\alpha -\tilde s    \;,
    \label{cons2} 
\end{align}
gauged  by the fields $e, a, \tilde a$, i.e. the einbein and the two independent worldline $U(1)$ gauge fields, respectively,
while  the constants $s$ and $\tilde s$ are suitable  Chern-Simons couplings. 
The color variables can be taken to be either commuting or anticommuting.
Here we choose the first option, which as a bonus may allow to study a classical limit 
of the color charge, as described by  Wong's equations \cite{Wong:1970fu}.
The kinetic term defines the phase space symplectic form,  leading to the 
Poisson brackets. The nonvanishing Poisson brackets are given by
 $\{x^\mu, p_\nu\}=\delta^\mu_\nu$, 
$\{c^a, \bar c_b\}=-\ii\delta^a_b$, $\{d^\alpha, \bar d_\beta\}=-\ii\delta^\alpha_\beta$, 
and used to verify that the constraints are indeed first class
\be
\{H, J\}=\{H, \tilde J\}=\{J, \tilde J\}= 0 \;.
\ee
Thus, the constraints can be gauged consistently.
Notice also the following Poisson brackets
\be
\{C^a, C^b\}= f^{abc} C^c \;,  \quad 
\{\tilde C^\alpha, \tilde C^\beta\}= \tilde f^{\alpha\beta\gamma} \tilde C^\gamma \;,  \quad 
\{C^a, \tilde C^\alpha\}= 0 \;,  \quad 
\{J, C^a\}= 0 \;, \ \  \text{etc}. 
\ee
The precise value of the Chern-Simons couplings $s$ and $\tilde s$
will be specified after transition to the quantum theory, 
which we discuss next using canonical quantization.

 The quantum theory of the particle has an associated Hilbert space  of ``wave-functions'' 
 on which the fundamental quantum operators act. These operators are the particle position and momentum,
 $\hat x^\mu, \hat p_\mu$, and two pairs of color variables, 
$\hat c^a, \hat c^\dagger_a $ and $\hat d^\alpha, \hat d^\dagger_\alpha $. 
The latter act as sets of creation and annihilation operators. 
Considering only the first set of color variables, $\hat c^a$, $\hat c^\dagger_a $,
they satisfy 
\be 
[\hat c^a,  \hat c^\dagger_b] =\delta^a_b, \quad  [\hat c^a,  \hat c^b] = 0 =[\hat c^\dagger_a,  \hat c^\dagger_b] 
\ee
with indices running up to $N_{A} = {\rm dim}\, G$, the dimension of the adjoint representation. 
They are naturally represented 
by ${\hat\cc}^{\dagger}_a \sim  \bar \cc_a$ and ${\hat\cc}^a \sim \partial/\partial \bar \cc_a$ 
 when acting on wave-functions of the form $\phi (x, \bar \cc)$. 
 The Taylor expansion of the latter reads 
\begin{align}
 \phi (x,  \bar \cc)= \phi(x)+ \phi^{a} (x) \bar\cc_a+\frac{1}{2!} \phi^{a b} (x) \bar\cc_a \bar\cc_b+ \dots,
\end{align}
and exposes sectors with different occupation numbers for the color variables, as measured by the 
number operator $\hat N={\hat\cc}^{\dagger}_a {\hat\cc}^a  \sim \bar \cc^a \partial/\partial \bar \cc^a$.
If one is interested only on wave-functions transforming in the adjoint, and not on tensor products of the adjoint,
one must impose a constraint that fixes the occupation number to 1, i.e., 
\be
(\hat N-1)  \phi (x,  \bar \cc) =  
\Big (\bar \cc^a \frac{\partial}{\partial \bar \cc^a} -1\Big ) \phi (x,  \bar \cc) = 0\;.
\ee
This is precisely the restriction that the quantum version of the first class constraint 
\eqref{cons1} should impose for meeting our purposes.
The quantization of this constraint must resolve ordering ambiguities, which we fix as in the harmonic oscillator 
by requiring a symmetric ordering
\be
J= \bar \cc_a \cc^a -s 
=  \frac12 (\bar \cc_a \cc^a + \cc^a \bar \cc_a)-s   \;,
\ee
leading to 
\be
\hat J= \frac12 (\hat c^\dagger_a \hat c^a +\hat c^a \hat c^\dagger_a) -s =
\hat c^\dagger_a \hat c^a + \frac{N_A}{2} -s =
\bar \cc^a \frac{\partial}{\partial \bar \cc^a} + \frac{N_A}{2} -s \;.
\ee
It is now simple to fix the Chern-Simons coupling $s$  to achieve occupation number 1
\be
\frac{N_A}{2} -s = -1   \qquad \to \qquad s= 1+ \frac{N_A}{2}\;.
\label{CS-coup}
\ee
A similar coupling with $\tilde s= 1+ \frac{\tilde N_A}{2}$
must be used for the other color sector as well, so to have physical wave-functions
transforming in the bi-adjoint, as desired. 
Thus, the particle wave-function  $\phi (x,  \bar \cc, \bar d) $ that satisfies the quantum constraints 
corresponding to $J$ and $ {\tilde J}$ has the form
\be
\phi (x,  \bar \cc, \bar d) =  \phi^{a\alpha} (x) \bar\cc_a \bar d_\alpha
\label{bi-adj}
\ee
and describes a   wave-function carrying indices of the bi-adjoint.
The remaining quantum constraint is the one corresponding  to the Hamiltonian $H$ in \eqref{ham}, 
whose operatorial version  takes the form
\be
\hat H = 
\frac12 \Big ( -\partial^2 + m^2 + y \Phi^{a\alpha}(x) f^{abc} \bar c^b \frac{\partial}{\partial \bar c^c}\,  
\tilde f^{\alpha\beta\gamma} \bar d^\beta  \frac{\partial}{\partial \bar d^\gamma} \Big ) \;.
 \ee
The constraint equation $\hat H  \phi (x,  \bar \cc, \bar d)  =0$ 
reproduces precisely the equations of  motion of the bi-adjoint field obtained from the Lagrangian \eqref{quad-act}.
Note that simple Lie groups have traceless generators, so that in this case 
the ordering of the color operators in the quantum version
 of  $C^a$ and $\tilde C^\alpha$ is inconsequential.
 We have thus completed the derivation of the worldline model for the bi-adjoint field theory.

From the  phase-space particle action \eqref{action1},  one may pass to configuration 
space (subscript `$\text{cs}$')
 by eliminating the momenta, finding
\be
S_{\text{cs}} =\int \dd\tau 
\Big [ \frac12 e^{-1}\dot x^2 + \ii \bar \cc_a (\partial_\tau +\ii a) \cc^a  + \ii \bar d_\alpha (\partial_\tau +\ii \tilde a) d^\alpha  
-\frac{e}{2} (m^2 -y C^a  \Phi^{a\alpha}(x) {\tilde C}^{\alpha})
 +s a +   {\tilde s}  {\tilde a}   \Big ]  
\label{action}
\ee
with $C^a$ and $\tilde C^\alpha$ already defined in \eqref{ham}. In the following, we are going to use an Euclidean version of this action, obtained by a Wick 
rotation\footnote{This is achieved  by sending  $\tau \to -\ii \tau$, 
accompanied by  $x^0\to -\ii x^4$ as well as  $a\to \ii a$ and $\tilde a\to \ii \tilde a$ for the worldline $U(1)$ gauge fields.}
($\ii S_{cs}\to - S_E$) producing
\be
S_E =\int \dd\tau 
\Big [ \frac12 e^{-1}\dot x^2 + \bar \cc_a (\partial_\tau +\ii a) \cc^a  +  \bar d_\alpha (\partial_\tau +\ii \tilde a) d^\alpha  
+\frac{e}{2} (m^2 -y C^a  \Phi^{a\alpha}(x) {\tilde C}^{\alpha})
 -\ii s a -\ii \tilde s  {\tilde a}) \Big ]  \;.
\label{eu-action}
\ee 

A covariant gauge fixing can be imposed by setting 
$(e(\tau),a(\tau),\tilde a(\tau))= (2T, \theta, \phi)$, with $T$ identified with the  usual Fock-Schwinger proper time, 
and $(\theta,\phi)$ two additional moduli, corresponding to two angles, related to the projection 
on occupation number 1 for each color sector. This gauge fixing is valid both for worldlines with the topology of 
a circle, as appropriate for the one-loop effective action, and for 
worldlines with the topology of an interval,  as appropriate for the propagator.
The difference is just on the measure on the moduli space. 
The gauge-fixed Euclidean action simplifies further. We now drop the subscript `E' and consider the parametrization 
of the whole worldline with $\tau\in [0,1]$, so that 
\be
S = T m^2  -\ii s \theta -\ii \tilde s  \phi +
\int_0^1\! \dd\tau 
\Big [ \frac{1}{4 T} \dot x^2 + \bar \cc_a (\partial_\tau +\ii \theta) \cc^a  +  \bar d_\alpha (\partial_\tau +\ii \phi) d^\alpha  
- T y C^a  \Phi^{a\alpha}(x) {\tilde C}^{\alpha}  \Big ]  \;,
\label{eu-gf-action}
\ee 
which appears in the path integral as $e^{-S}$. 

Let us comment on the vertex operators it produces.
The background $\Phi^{a\alpha}(x)$ can be taken to be  a plane wave  with fixed colors $a',\alpha'$ 
and incoming momentum $k^\mu$
\be
\Phi^{a\alpha}(x) = \delta^{a a'}  \delta^{\alpha \alpha'}  e^{\ii k \cdot x}\;.
\ee
Treating it as a perturbation, one may expand the exponential of the interaction part $e^{-S_{\text{int}}}$ on which it sits.
Then, the leading term of the expansion identifies the vertex operator
\be
 y T \int_0^1 \! \dd\tau\, C^{a'}(\tau) {\tilde C}^{\alpha'} (\tau) e^{\ii k \cdot x(\tau)} = y T V^{(0)}[k, a',\alpha'] \;,
\label{vo}
\ee
which reproduces the one we conjectured  earlier on in \eqref{bi-adjoint-vertex}, with the correct power of proper time $T$
and coupling constant $y$, which must go along with it. 

\section{Effective action}
\label{effective-action}
We have now the required tools to quantize the theory using path integrals. We wish to find a worldline representation of the one-loop effective action  $\Gamma[\Phi]$ of the bi-adjoint particle. As we did in the previous Section, 
we first consider 
 the method with the matrix-valued action and path ordering, and then the one with the auxiliary color  variables. 
\subsection{  The matrix-valued approach}

As already discussed in sec. \ref{Wmva},
the matrix-valued action requires a path ordering prescription
to define properly the corresponding path integral.
The Schwinger proper-time representation of the one-loop effective action
of the bi-adjoint field has already been anticipated in 
\eqref{pi-new1}, where the trace of the heat kernel is computed by the path integral with
periodic boundary conditions.
The periodic paths $x^\mu(\tau)$ with $x^\mu(0)=x^\mu(1)$ can be split as
  \be 
 x^\mu(\tau)=x_0^\mu+q^\mu(\tau) \;,
 \label{3-split}
 \ee
  where $x_0^\mu$  is any point on the loop traced by $x^\mu(\tau)$, while
  $q^\mu(\tau)$ are the remaining fluctuations that must satisfy
vanishing Dirichlet boundary conditions  (i.e. $q^\mu(0)=q^\mu(1)=0$).
Then, one integrates over $x^\mu_0$ and $q^\mu(\tau)$ separately,
and the effective action takes the form
\begin{align}
\Gamma[\Phi] = -\frac12 
\int_0^\infty \frac{\dd T}{T} e^{-m^2T}
\int d^D x_0\
{\rm tr} \int_{\cal D} Dq \, 
{\bf T}\, e^{-S[x]} \;,
\label{pi-new3}
\end{align}
where $S[x]$ is the action in \eqref{2.14}, and
$\cal D$ indicates the Dirichlet boundary conditions on $q^\mu(\tau)$  mentioned above.
Other methods for extracting the constant $x_0$
are possible, and will be discussed in the next section.

The effective action \eqref{pi-new3} may now be computed perturbatively 
in terms of the Wick contractions of the $q$-propagators (see appendix \ref{appA} for further details)
\be 
\Gamma[\Phi] = \int d^D x_0\left[
-\frac12 
\int_0^\infty \frac{\dd T}{T}  \frac{e^{-m^2T}}{\left(4\pi T\right)^\frac{D}{2}} 
\, {\rm tr}\,
\la {\bf T}\, e^{-S_{\text{int}}} \ra
\right ]
\label{pi-new2}
\ee 
with 
\be 
S_{\text{int}}  =  T y \int_0^1 \!  \dd\tau\,  \hat \Phi(x(\tau))  \;.
\label{3-int}
\ee
The perturbative computation gives an answer of the form 
\be
 {\rm tr}\, 
 \la {\bf T} e^{-S_{\text{int}}} \ra = 
   \sum_{n=0}^{\infty} {\rm tr}\, a_n(x_0) T^n\;,
   \label{pert-exp}
 \ee
 where $a_n(x_0)$ are  the so-called Seeley-DeWitt (or  heat kernel) coefficients.
 Inserting this expansion in  \eqref{pi-new2}, and renaming $x_0\to x$, gives 
\ba
\Gamma[\Phi] &= \int d^D x \left [
 - \frac 12 \sum_{n=0}^{\infty} \frac{{\rm tr}\, a_n(x)}{(4\pi)^{\frac{D}{2}}}  
 \int_0^\infty \frac{dT}{T}  e^{-m^2 T} 
   T^{n -\frac{D}{2}}
 \right ] 
 \\
 &= \int d^D x \left [
 - \frac 12 \sum_{n=0}^{\infty} \frac{{\rm tr}\, a_n(x)}{(4\pi)^{\frac{D}{2}}} 
 \frac{1}{(m^2)^{n -\frac{D}{2}}}
 \int_0^\infty \frac{dT}{T}  e^{-T} 
   T^{n -\frac{D}{2}}
 \right ] 
  \\
 &= \int d^D x \left [
 - \frac 12 \sum_{n=0}^{\infty} \frac{{\rm tr}\, a_n(x)}{(4\pi)^{\frac{D}{2}}} 
 \frac{\Gamma[ n -\tfrac{D}{2}]}{(m^2)^{n -\frac{D}{2}}} 
 \right ] \;.
 \label{eff-act}
\ea
We have integrated the proper time to obtain the usual gamma function.
At $D=6$, we see divergences for $n=0,1,2,3$ in the effective action, as captured by the poles of the gamma function.
The one for $n=3$ is mass independent and corresponds to the usual logarithmic divergence seen in dimensional regularization. The specific calculation  of the coefficients in \eqref{pert-exp} 
 can be done as in \cite{Bastianelli:1992ct, Fliegner:1993wh, Fliegner:1997rk,Bastianelli:2020fdi},
 and gives the answer 
\be
a_0 = 1 \;, \quad 
a_1  = - y \hat \Phi \;, \quad 
a_2 = \frac{y^2}{2} \hat \Phi^2   \;, \quad 
a_3 =  -\frac{y^2}{6}   \hat \Phi^3 + \frac{y^2}{12} \hat \Phi \partial^2 \hat \Phi \;,
\label{coeffs-SdW}
\ee
which we have simplified by adding total derivatives, and neglecting matrix orderings that  are inconsequential under the trace. The calculation of these coefficients is briefly outlined in appendix \ref{appA}.

 \subsubsection{Application: the beta function at $D=6$}  
 The $n=3$ pole in the effective Lagrangian at $D=6 -\epsilon$ is 
\be
{\cal L}_{\text{eff, div}} =
 - \frac 12 \frac{ {\rm tr}\, a_3(x)}{(4\pi)^3} \Gamma[\tfrac{\epsilon}{2}] 
 =  -  \frac{{\rm tr}\, a_3(x)}{(4\pi)^3}  \frac{1}{\epsilon} \;,
\label{ea-div}
   \ee
Then, using \eqref{coeffs-SdW} the diverging part of the effective Lagrangian \eqref{ea-div} becomes
\be
{\cal L}_{\text{eff, div}}  =    \frac{1}{(4\pi)^3}  \frac{1}{\epsilon} \left(\frac{y^3}{6}  {\rm tr}\,  \hat \Phi^3  
 - \frac{y^2}{12}  {\rm tr}\, \hat \Phi \partial^2 \hat \Phi  \right )\;.
  \label{div2}
   \ee
 Let us now relate the terms in \eqref{div2} to the one in \eqref{lag-phi3}, and 
 recognize the counterterms needed to renormalize the  theory at one-loop.
 We find
 \ba
 {\rm tr}\,  \hat \Phi^2 &= T(A) \tilde T(A) (\Phi^{a\alpha})^2
 \\
 {\rm tr}\,  \hat \Phi^3 &= \frac14 T(A) \tilde T(A) f^{abc} \tilde f^{\alpha \beta \gamma} \Phi^{a \alpha} \Phi^{b \beta} \Phi^{c \gamma} \;,
 \label{cal}
  \ea
  where $T(A)$ indicates the index in the adjoint representation\footnote{
  The index $T(R)$ of a representation $R$ is defined by
 ${\rm tr} (T_{R}^aT_{R}^b)= T(R) \, \delta^{ab} $. It coincides with its quadratic Casimir if $R$ is the adjoint representation. In particular,
 one has $T(A)=C_2(A)=N$ for the adjoint representation of $SU(N)$.
 }.
      These divergences ask for wave function and coupling constant renormalizations, 
  obtained by adding to  \eqref{lag-phi3} the counterterms
   \be
   {\cal L}_{\text{ct}}= (Z_\Phi-1) \tfrac12 \Phi^{a\alpha}  (-\partial^2) \Phi^{a\alpha} 
   +
   (Z_y-1) \tfrac{y}{3!}  f^{abc} \tilde f^{\alpha \beta \gamma} 
   \Phi^{a \alpha} \Phi^{b \beta} \Phi^{c \gamma} 
      \ee
with 
\be
Z_\Phi-1 = - \frac16 T(A) \tilde T(A)
\frac{y^2}{(4\pi)^3} \frac{1}{\epsilon} \;, \qquad 
Z_y-1 = -\frac14  T(A) \tilde T(A)
 \frac{y^2}{(4\pi)^3} \frac{1}{\epsilon} \;.
\ee  
 These counterterms produce a vanishing one-loop beta function,  
as recently  discovered  in \cite{Gracey:2020baa}. 
In more details, parameterizing the counterterms with coefficients $a_1$ and $c_1$ as

 \ba
 Z_\Phi-1 &= a_1  \frac{y^2}{(4\pi)^3} \frac{1}{\epsilon}
  \;,  \qquad  a_1= -\frac16 T(A) \tilde T(A)\;,
  \\
    Z_y-1 &= c_1 \frac{y^2}{(4\pi)^3} \frac{1}{\epsilon} \;, \qquad   c_1= -\frac14 T(A) \tilde T(A)\; ,
\label{25}
 \ea
the one-loop beta function is computed by  
 \be
 \beta(y) = \left (c_1- \frac32 a_1\right ) \frac{y^3}{(4\pi)^3} = 0  \;.
 \label{26}
  \ee
 In \cite{Gracey:2020baa} the calculation of the beta function has been pushed to four loops 
  (and actually to five loops in \cite{Borinsky:2021jdb}),  with the two-loop result 
    indicating the asymptotic freedom of the theory. The latter  would furnish a nice test on the 
structure of worldline methods at higher loops \cite{Sato:1998kq, Schubert:2001he}.

There is also a mass renormalization that we have ignored so far. To complete the one-loop renormalization 
of the theory, let us  identify the counterterm needed for renormalizing the mass. To achieve that, 
  we have to treat the mass perturbatively in  \eqref{pi-new2}. Expanding the term $e^{-m^2 T}$ 
we find (up to total derivatives and up to a constant that renormalizes the vacuum energy)
 an extra contribution to ${\rm tr}\, a_3(x)$ of the form
 \be
 \Delta {\rm tr}\, a_3(x) =-\frac12 m^2 y^2 {\rm tr}\,  \hat \Phi^2  =
 -\frac12 m^2 y^2   T(A) \tilde T(A) (\Phi^{a\alpha})^2
  \ee
 giving an extra divergence of the effective Lagrangian
 \be
\Delta {\cal L}_{\text{eff, div}} = 
 -  \frac{\Delta {\rm tr}\, a_3(x)}{(4\pi)^3} \frac{1}{\epsilon}
  =    \frac{1}{(4\pi)^3}  \frac{1}{\epsilon} 
  \left(\frac{y^2 m^2}{2}  T(A) \tilde T(A) (\Phi^{a\alpha})^2
   \right ) \;.
  \label{extra-div}
   \ee
This is canceled by adding to \eqref{lag-phi3} the additional counterterm
  \be
\Delta {\cal L}_{\text{ct}}= (Z_{m}-1) \tfrac12 m^2 (\Phi^{a\alpha})^2 
        \ee
with 
\be
Z_{m}-1 = - T(A) \tilde T(A) \frac{y^2}{(4\pi)^3} \frac{1}{\epsilon} \;,
\ee  
which leads to the mass anomalous dimension $\gamma_m$ (defined as usual by
$\gamma_m =\frac{1}{m} \frac{d m}{d \ln \mu}$)  
\be
\gamma_m(y) = -\frac{5}{12} T(A) \tilde T(A) \frac{y^2}{(4\pi)^3} \;.
\ee
 
\subsection{ The color variables approach}
Let us now proceed with the auxiliary color variables approach. The general structure 
 of the one-loop effective action was anticipated in \eqref{schematic-action},  with the role of $\Psi^{(0)}$  now taken by $\Phi^{a\alpha}$. Our auxiliary variables $\chi$  are the einbein $e$, the gauge fields $a$, $\tilde a$, and the 
color variables $c$, $\bar c$, $d$, $\bar d$.  Hence, the one-loop effective action induced by a self-coupled bi-adjoint scalar particle is 
\begin{equation}
 \Gamma[\Phi]= \int_{S^1} \frac{D x De Da D\tilde a D c D \bar c Dd D\bar d }{\text{Vol}(\text{Gauge})}\,  e^{-S_E}
\end{equation}
with $S_E$ given in eq.~\eqref{eu-action}, with $\tau \in [0,1]$ describing the circle.
To bring it in a computable form, we must gauge-fix it. Let us review the main steps adapted to the circle $S^1$. 

As anticipated, on the circle one may fix the worldline  gauge fields to constants \linebreak $(e(\tau), a(\tau), \tilde a (\tau))=
(2T, \theta, \phi)$, with the latter playing the role of moduli, i.e. gauge invariant configurations that must be integrated over 
a suitable moduli space \cite{Bastianelli:2005vk}.
The measure on the moduli space is given by Faddeev-Popov determinants, which are constant in our case 
(i.e. they do not depend on the moduli), except that there is a 
factor $\frac{1}{T}$ that takes into account the symmetry generated by constant translations on the circle: 
this factor avoids relative overcountings of paths with different proper time $T$.
Then, fixing the overall normalization to match that of a real scalar, we have 
\begin{align}
\Gamma[\Phi] = -\frac12 
\int_0^\infty \frac{\dd T}{T} 
\int_{0}^{2\pi} \frac{\dd \theta}{2\pi}  \int_{0}^{2\pi}\frac{\dd \phi}{2\pi}  
\int_{\cal P} D x D c D \bar c Dd D\bar d \   e^{-S} \;,
\label{pi-1}
\end{align}
where the gauge-fixed action $S$ is given in \eqref{eu-gf-action},
and $\cal P$ denotes periodic boundary conditions for the remaining path integration variables.
The integration over the moduli space is set to cover the moduli space only once.
The remaining path integral is normalized to that of a free particle, arising
when the interactions  are set to vanish (more on this later). 

Let us further manipulate the remaining path integral to unearth the zero modes that may be present on the circle.
Let us first address the zero modes of the coordinates $x^\mu$. They are present when 
the potential is treated perturbatively. 
These zero modes satisfy $\partial_\tau^2\, x^\mu(\tau) =0$ and are therefore constant configurations.
 Their separation is generically achieved by splitting  
 \be 
 x^\mu(\tau)=x_0^\mu+q^\mu(\tau) \;,
 \ee
  where $x_0^\mu$ are the constant zero modes, and 
$q^\mu(\tau)$ are the remaining quantum fluctuations without the zero modes. 
There is a variety of ways of achieving this split, with the most common ones corresponding to 
setting Dirichlet boundary conditions  
on  $q^\mu(\tau)$  (i.e. imposing $q(0)=q(1)=0$) or using the ``string-inspired'' boundary conditions
(that is requiring $\int_0^1\dd \tau\, q^\mu(\tau)=0$). They give equivalent results, see \cite{Bastianelli:2020fdi}
for a recent application and comparison of the two methods.
The integration of the zero modes factorizes, as perturbatively there is no action for them,
and there remains the path integral over the quantum fluctuations $q^\mu$.
At this stage, we may extract the zero modes $x_0$ to rewrite the path integral \eqref{pi-1} as
\begin{align}
\Gamma[\Phi] = -\frac12 
\int_0^\infty \frac{\dd T}{T} 
\int_{0}^{2\pi} \frac{\dd \theta}{2\pi}  \int_{0}^{2\pi}\frac{\dd \phi}{2\pi}  
\int d^D x_0
\int_{\bar {\cal P}} Dq 
\int_{\cal P} D c D \bar c Dd D\bar d \   e^{-S} \;,
\label{pi-2}
\end{align}
where now $\bar{\cal P}$ denotes periodic boundary conditions with a constraint that eliminates the zero modes
(it depends on the chosen method). We have considered arbitrary dimensions $D$, 
also in view of the application of dimensional regularization, though the case $D=6$ is the most interesting one
as in such dimensions the model is renormalizable. 

Let us now address the color variables $c$ and $d$. For non-vanishing constant gauge fields $\theta$ and $\phi$,
and periodic boundary conditions, the color variables do not have any zero mode. 
Indeed, $\theta$ and $\phi$ act as a kind of mass term,
i.e. a Wick rotated frequency-squared term for these harmonic-like oscillators, see their action in \eqref{eu-gf-action}.
Zero modes appear only for  $\theta=0$ and $\phi=0$, and we will consider their effect later on.
It is also possible to remove the couplings to  $\theta$ and $\phi$ by shifting those couplings  to the boundary conditions
by field redefinitions
\begin{align}
 c(\tau) \to   e^{-\ii \theta \tau} c(\tau), \ \  \bar c(\tau) \to   e^{\ii \theta \tau} \bar c(\tau) \;, 
\qquad 
 d(\tau) \to   e^{-\ii \phi \tau} d(\tau), \ \ \bar d(\tau) \to   e^{\ii \phi \tau} \bar d(\tau) \;,
\end{align}
 which implies that the new fields $c$ and $d$ thus obtained satisfy twisted boundary conditions (which we denote by $\cal{T}$)
\begin{align}
 c(1)=e^{i\theta}c(0), \quad \bar{c}(1)=e^{-i\theta}\bar{c}(0)\,,  \qquad
  d(1)=e^{i\phi}d(0), \quad \bar{d}(1)=e^{-i\phi}\bar{d}(0) \;.
\end{align}
The effective action then takes the form
\begin{align}
\Gamma[\Phi] = -\frac12 
\int_0^\infty \frac{\dd T}{T}  e^{-m^2T} \!\!
\int_{0}^{2\pi} \frac{\dd \theta}{2\pi}  e^{\ii s\theta} \!\!
\int_{0}^{2\pi}\frac{\dd \phi}{2\pi}  e^{\ii \tilde s \phi} \!\!
\int d^D x_0
\int_{\bar {\cal P}} Dq 
\int_{\cal T} D c D \bar c Dd D\bar d \   e^{-S_0}, 
\label{pi-3}
\end{align}
where the worldline action $S_0$ now reads 
\begin{equation} 
\begin{aligned}
 S_0=
\int_0^1 \dd\tau 
\Big [ \frac{1}{4 T} \dot q^2 + \bar \cc_a \dot  \cc^a  +  \bar d_\alpha \dot  d^\alpha  
- T y C^a  \Phi^{a\alpha}(x) {\tilde C}^{\alpha}  \Big ]   
\end{aligned} 
\end{equation}
since we have extracted the mass term $e^{-m^2T}$ and the Chern-Simons couplings $e^{\ii s\theta+ \ii \tilde s \phi}$.

From this expression we see that in a perturbative expansion we may treat the potential as a perturbation, while
recognizing from the kinetic term the free propagators of the particle coordinates $q^\mu$ and auxiliary color variables
$c,\bar c, d, \bar d$.
Extracting the normalization due to the free path integral we find
\be 
\Gamma[\Phi] = \int d^D x_0\left[
-\frac12 
\int_0^\infty \frac{\dd T}{T}  \frac{e^{-m^2T}}{\left(4\pi T\right)^\frac{D}{2}}
\int_{0}^{2\pi} \frac{\dd \theta}{2\pi}  
\frac{e^{\ii s\theta}}{\left(2\ii\sin\frac{\theta}{2}\right)^{N_{A}}}
\int_{0}^{2\pi}\frac{\dd \phi}{2\pi} 
\frac{e^{\ii \tilde s\phi}}{\left(2\ii\sin\frac{\phi}{2}\right)^{\tilde N_{A}}}
\la e^{-S_{\text{int}}} \ra
\right ]
\label{pi-4}
\ee 
with the term in square bracket representing the QFT one-loop effective Lagrangian ${\cal L}_{\text{eff}}$
(that, of course, must be renormalized, as for example by treating the arbitrary dimension $D$ as in dimensional regularization).
The perturbation is here given by
\be \label{int-action}
S_{\text{int}}  = - T y \int_0^1 \!  \dd\tau\,  C^a  \Phi^{a\alpha}(x) {\tilde C}^{\alpha}  \;.
\ee
 The free propagators that go along with this representation, and 
 needed to compute the average $ \la e^{-S_{\text{int}}} \ra$
 by Wick contractions, take into account the boundary conditions. They are given by
\be \begin{aligned}
\la q^{\mu}(\tau)q^{\nu}(\sigma)\ra &= -2T \delta^{\mu\nu}\del(\tau,\sigma) \;, 
\\
\la c^{a}(\tau)\bar{c}_{b}(\sigma)\ra &= \delta^{a}_{b}\del_{_{\cal{T}}}(\tau,\sigma;\theta)
\; ,
\\
\la d^{\alpha}(\tau)\bar{d}_{\beta}(\sigma)\ra &= \delta^{\alpha}_{\beta}\del_{_{\cal{T}}}(\tau,\sigma;\phi) \;,
\end{aligned}
\label{prop}
\ee
where\footnote{We use the string inspired method to factor out the zero modes from $q$. The constant term $-\frac{1}{12}$ 
could also be dropped from the propagator $\Delta(\tau,\sigma)$  \cite{Schubert:2001he}.
As for the propagator $\del_{_{\cal{T}}}(\tau,\sigma;\theta)$, one may check that it satisfies the Green's equation 
$\partial_\tau \del_{_{\cal{T}}}(\tau,\sigma;\theta) =\delta(\tau-\sigma)$, and the twisted boundary conditions 
$\del_{_{\cal{T}}}(1,\sigma;\theta) = e^{\ii \theta} \del_{_{\cal{T}}}(0,\sigma;\theta)$
and  $\del_{_{\cal{T}}}(\tau,1;\theta) = e^{-\ii \theta} \del_{_{\cal{T}}}(\tau,0;\theta)$.
}
\be \begin{aligned}
\del(\tau,\sigma)& =\frac{1}{2}|\tau-\sigma|-\frac{1}{2}(\tau-\sigma)^{2}-\frac{1}{12}\;,\\
\del_{_{\cal{T}}}(\tau,\sigma;\theta) &= 
\frac{1}{2\ii\sin\frac{\theta}{2}} \left[ e^{\ii \frac{\theta}{2}} \Theta(\tau-\sigma) 
+e^{-\ii \frac{\theta}{2}} \Theta(\sigma-\tau) 
\right]\;, 
\end{aligned}\ee
where $\Theta(x)$ is the standard Heaviside step-function with $\Theta(0)=\frac12$. 

Eq.~\eqref{pi-4} is our final form of the worldline representation of the one-loop effective action
of the bi-adjoint scalar,  with the additional color variables. It is 
 readily  calculable in perturbation theory. One may compare it with \eqref{pi-new2}, which used instead a matrix valued action.

\subsubsection{ Application: the self-energy}
\label{Self-energy}
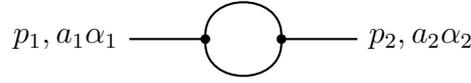
\begin{figure}[h]
\centering
\begin{tikzpicture}[thick, transform shape]
\begin{feynman}[small]
\vertex (c0);
\vertex[right = 3 of c0] (c1);
\vertex[left = 3 of c1] (c2){\(p_1, a_1 \alpha_1\)};
\vertex[right = 3 of c0] (c3){\(p_2, a_2 \alpha_2\)};
\diagram*{
a -- b[dot]-- [ half left] c[dot]
-- [ half left] b,
(c) --  (c1),
};
\end{feynman}
\end{tikzpicture}
\caption{Self-energy contribution to the bi-adjoint propagator at one-loop.}
\label{fig:self-energy}
\end{figure}

Let us now consider a simple application of \eqref{pi-4},  the calculation of the self-energy  
of the bi-adjoint scalar, as described  by Fig.~\ref{fig:self-energy}.

To start with, let us first check the normalization of the effective action  \eqref{pi-4}, by setting $S_{\text{int}}$ to vanish,
 and verifying that it contains the correct number of degrees of freedom circulating in the loop. To do that, and to perform in a more 
convenient way the integration over the angular moduli, we change variables from 
 $\theta$ and $\phi$ to the Wilson loop variables  $z= e^{-\ii \theta}= e^{-\ii \int_0^1 \dd \tau\, a(\tau) } $
  and $w= e^{-\ii \phi}= e^{-\ii \int_0^1 \dd \tau\, \tilde a(\tau) } $, so that the original integration
 is mapped to a contour integration, the unit circle $\gamma$ of the complex plane for each variable, e.g. for $\theta$
\be
        \int_0^{2\pi} \frac{d\theta}{2\pi} = \varointctrclockwise_\gamma \frac{d z}{2\pi \ii z}
  \;, \quad  
   2 \ii \sin\tfrac{\theta}{2} = \frac{1-z}{\sqrt{z}}
   \;, \quad  
      \int_{0}^{2\pi} \frac{\dd \theta}{2\pi}  
\frac{e^{\ii s\theta}}{\left(2\ii\sin\frac{\theta}{2}\right)^{N_{A}}}= 
\varointctrclockwise_\gamma \frac{\dd z}{2\pi \ii } \frac{1}{z^2} \frac{1}{(1-z)^{N_A}} \;,
\label{65}
        \ee
where we used the value of the Chern-Simons coupling $s=1+\frac{N_A}{2}$.
One may notice that at $\theta =0$ 
the auxiliary color variables $c, \bar c$ develop  zero modes, which correspond to  
the pole at $z=1$. 
This divergence is due to the determinant of the free path integral that vanishes at such a point because of the zero modes.
Previously we postponed the discussion of these zero modes, but now we see 
their effect and the way we should treat them. They cause a singularity on the integration contour at $z=1$, which 
we push out of the integration contour (indicated by $\gamma_-$, see Fig.~\ref{contour1}). 
This prescription gives the expected answer. With it, only the poles at $z=0$ 
are responsible for the projection and we get 
\be
\varointctrclockwise_{\gamma_-} \frac{\dd z}{2\pi \ii } \frac{1}{z^2} \frac{1}{(1-z)^{N_A}} = \frac{\dd }{\dd z } (1-z)^{-N_A}\Big |_{z=0} = N_A \;,
        \ee
which is the expected number.

Note that inclusion of the pole $z=1$ inside the contour would give a total vanishing result,
which means that the previous answer can be obtained also by viewing the complex plane as the Riemann sphere, 
and considering only the poles outside the contour  $\gamma_-$.
This fact is related to a redefinition of the Wilson loop variable 
$z\to z' = \frac{1}{z} = e^{\ii \theta}$, (or equivalently to a different gauge fixing for $a(\tau)$, namely $\theta\to -\theta$),
that must give equivalent results. This change of variables maps the contour   $\gamma_-$ of the $z$-plane to the contour $\gamma_+$
of the $z'$-plane, which now includes the pole at $z'=1$, see Fig.~\ref{contour2}
(a change of sign has been taken care of by reversing the orientation of the contour).
\begin{figure}[ht]
 \centering
 \begin{minipage}[b]{0.45\linewidth}
       \begin{center}
\begin{tikzpicture}[decoration={%
   markings,%
   mark=at position 2.5cm with {\arrow[black]{stealth};}}]
 \draw[->] (-1.5,0) -- (1.5,0)  node[anchor=north west] {Re $(z)$};  
 \draw[->] (0,-1.5) -- (0,1.5) node[anchor=south east] {Im $(z)$};
\filldraw [black] (1,0) circle [radius=.6pt] ;
\draw[thick,postaction=decorate] ([shift=(5:1cm)]0,0) arc (5:355:1cm);
\draw[thick,postaction=decorate] ([shift=(90:.1cm)]1,0) arc (90:270:.1cm);
\end{tikzpicture}
          \end{center}
   \caption{Integration contour $\gamma_-$ in  \\ $z$-plane.
    Pole at  $z=1$ is excluded.}
   \label{contour1}
 \end{minipage}
 \quad
 \begin{minipage}[b]{0.45\linewidth}
        \begin{center}
\begin{tikzpicture}[decoration={%
   markings,%
   mark=at position 2.5cm with {\arrow[black]{{stealth}};}}]
  \draw[->] (-1.5,0) -- (1.5,0)  node[anchor=north west] {Re $(z')$};  
 \draw[->] (0,-1.5) -- (0,1.5) node[anchor=south east] {Im $(z')$};
\filldraw [black] (1,0) circle [radius=.6pt] ;
\draw[thick, postaction=decorate] ([shift=(5:1cm)]0,0) arc (5:355:1cm);
\draw[thick] ([shift=(270:.1cm)]1,0) arc (270:360:.1cm);
\draw[thick] ([shift=(0:.1cm)]1,0) arc (0:90:.1cm);
\end{tikzpicture}
       \end{center}
  \caption{Integration contour $\gamma_+$ in  \\ $z'$-plane.
    Pole at   $z'=1$ is included.}
  \label{contour2}
\end{minipage}
\end{figure}

Then eq.~\eqref{pi-4} reduces to 
\begin{align}
\Gamma[\Phi] = -\frac12 
\int d^D x_0
\int_0^\infty \frac{\dd T}{T}  \frac{e^{-m^2T}}{\left(4\pi T\right)^\frac{D}{2}}
\left (
N_A \tilde N_A +\dots
\right ) \;,
\label{pi-5}
\end{align}
which reproduces the expected degrees of freedom of the bi-adjoint scalar\footnote{A factor of 1 instead of 
$N_A \tilde N_A$ corresponds to a free real scalar.}.

After this check, we are ready to study the self-energy of Fig.~\ref{fig:self-energy}.
For that we consider a background given by the sum of two-plane waves
with quantum numbers $a_1,\alpha_1, p_1$ and $a_2,\alpha_2, p_2$ 
\be 
\Phi^{a\alpha}(x) = \delta^{a a_1}  \delta^{\alpha \alpha_1}  e^{\ii p_1 \cdot x(\tau)}
+ \delta^{a a_2}  \delta^{\alpha \alpha_2}  e^{\ii p_2 \cdot x(\tau)} \;.
\label{back-2-pw}
\ee
Expanding $S_{\text{int}}$ in \eqref{pi-4}
and keeping the contribution linear in each plane wave, we find the contribution of two vertex operators 
of the form given in \eqref{vo}
\be
\la e^{-S_{\text{int}}}\ra \ \to \ y^2 T^2 \la V(p_1,a_1,\alpha_1) V(p_2,a_2,\alpha_2)  \ra \;.
\ee
The integration over the zero modes produce a delta function for momentum conservation
and  one finds the self-energy correction
\be
\Gamma[\Phi] \ \to \  (2\pi)^{D}\delta^{D}(p_1+p_2) \Pi^{a_1\alpha_1, a_2\alpha_2}(p)\;,
\ee
where $p\equiv p_1=-p_2$ and 
\begin{align}
\Pi^{a_1\alpha_1, a_2\alpha_2}(p) = 
& -\frac{y^2}{2} 
\int_0^\infty \dd T  \frac{e^{-m^2T} T}{\left(4\pi T\right)^\frac{D}{2}}
\varointctrclockwise_{\gamma_-} \frac{\dd z}{2\pi \ii } \frac{z^{-2}}{(1-z)^{N_A}}
\varointctrclockwise_{\gamma_-} \frac{\dd w}{2\pi \ii } \frac{w^{-2}}{(1-w)^{\tilde N_A}}
\label{self-ernegy-ex}\\
& \quad \times   \la V_q(p,a_1,\alpha_1) V_q(-p,a_2,\alpha_2)  \ra \nonumber
\end{align}
with the vertex operators $V_q$ depending only on the $q$ coordinates without the zero modes.
Carrying out the Wick contractions, one finds
\ba 
&\la V_q(p,a_1,\alpha_1) V_q(-p,a_2,\alpha_2)  \ra \\
&= \int_{0}^{1}d\tau \int_{0}^{1}d\sigma \, 
  \la C^{a_{1}}(\tau)C^{a_{2}}(\sigma)\ra
   \la \tilde{C}^{\alpha_{1}}(\tau)\tilde{C}^{\alpha_{2}}(\sigma)\ra 
 \la e^{i p\cdot q(\tau)} \, e^{-i p\cdot q(\sigma)}\ra \\
 &=f^{a_{1} b c}f^{a_{2}bc} \tilde{f}^{\alpha_{1}\beta\gamma}\tilde{f}^{\alpha_{2}\beta\gamma} 
 \frac{z w}{(1-z)^{2}(1-w)^{2}} \int_{0}^{1}d\tau \int_{0}^{1}d\sigma \, e^{-2Tp^{2}\del_{0}(\tau,\sigma)}\\
 &=\delta^{a_{1}a_{2}}\delta^{\alpha_{1} \alpha_{2}} T(A) \tilde{T}(A)\frac{z w}{(1-z)^{2}(1-w)^{2}}
 \int_{0}^{1}d\tau  \, e^{-Tp^{2}(\tau-\tau^{2})}\;,
 \ea 
 where $\del_{0}(\tau,\sigma) = \del(\tau,\sigma)-\del(\tau,\tau)$, and  we used translational invariance of the string inspired propagator to get the last line.
Next, using that the complex integration gives one, we find that the proper time integration yields 
\be 
\Pi^{a_1\alpha_1, a_2\alpha_2}(p) =
-\frac{y^{2}}{2(4\pi)^{\frac{D}{2}}}
\delta^{a_{1} a_{2} } \delta^{\alpha_{1} \alpha_{2} }
T(A) \tilde T(A)
\, (P^{2})^{\frac{D}{2} -2 }\, \Gamma\left(2-\tfrac{D}{2}\right) \;,
\ee
where we defined 
\be 
(P^{2})^{x} \equiv \int_{0}^{1}d\tau \left( m^{2}+p^{2}(\tau-\tau^{2})\right)^{x}\;.
\label{74}
\ee
This is the unrenormalized contribution to the self-energy.
As a check, one may extract the UV divergences
related to wave-function and mass renormalizations using dimensional regularization,
which are easily seen to match the ones calculated in the previous Section.

\begin{figure}[htb]
\centering
\begin{tikzpicture}[thick, transform shape]
\begin{feynman}[small]
\vertex (c0);
\vertex[right = 3 of c0] (c1);
\vertex[left = 3 of c1] (c2){\(p_1, a_1 \alpha_1\)};
\vertex[right = 3 of c0] (c3){\(p_2, a_2 \alpha_2\)};
\diagram*{
a -- b[dot]-- [ half left, dashed ] c[dot]
-- [ half left, dashed] b,
(c) --  (c1),
};
\end{feynman}
\end{tikzpicture}
\caption{Self-energy contribution to the bi-adjoint propagator at one-loop due to a virtual scalar particle.
The charge of the particle in the loop is specified by occupation numbers $r$ and $\tilde r$. 
}
\label{fig:self-energy2}
\end{figure}
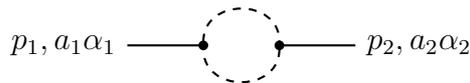

We have used the Chern-Simons couplings  to select occupation number 1 in each colored sector, thus making sure that there is precisely 
a bi-adjoint scalar particle circulating in the loop.
However, one could modify our worldline theory by selecting different occupation numbers, say $r$ and  $\tilde r$, for the color variables.
This corresponds to a differently charged particle circulating in the loop, but coupled to the same background bi-adjoint field
$\Phi^{a\alpha}$, see Fig.~\ref{fig:self-energy2}.

\noindent This is achieved by setting the Chern-Simons  couplings to 
\be
s= r+ \frac{N_A}{2}\;, \qquad \tilde s= \tilde r+ \frac{\tilde N_A}{2}\;, \qquad
\ee
as compared to our previous eq.~\eqref{CS-coup}. For example, setting $(r,\tilde r)=(0,0)$ gives an uncharged particle 
that should decouple from the loop,
setting $(r,\tilde r)=(2,0)$ would give a particle that transforms in the symmetric products of two adjoints for the group $G$ and scalar under 
$\tilde G$, which should also decouple, while
setting $(r,\tilde r)=(2,1)$ would give a particle that transforms in the symmetric products of two adjoints for the group $G$ and in the adjoint for 
$\tilde G$, which should give a nontrivial contribution to the self-energy, as depicted in Fig.~\ref{fig:self-energy2}.
Thus, using the above Chern-Simons  couplings, which modify eq.~\eqref{65},
 we find the following contribution to the self-energy
\be 
\Pi^{a_1\alpha_1, a_2\alpha_2}_{r\tilde r}
(p) =
-\frac{y^{2}_{r\tilde r}}{2(4\pi)^{\frac{D}{2}}}
\delta^{a_{1} a_{2} } \delta^{\alpha_{1} \alpha_{2} }
T(A) \tilde T(A) {N_A+r \choose N_A+1} {\tilde N_A+\tilde r \choose \tilde N_A+1} 
\, (P^{2})^{\frac{D}{2} -2 }\, \Gamma\left(2-\tfrac{D}{2}\right)\;,
\label{76}
\ee
where we have denoted by $y_{r\tilde r}$ the coupling constant of the new particle to the bi-adjoint field, and where 
the definition of $(P^{2})^x$ in eq.~\eqref{74}
should contain a different mass $m_{r\tilde r}$ instead of $m$.
 The binomial coefficients are defined to vanish for $r=0$  and $\tilde r=0$, as usual,
as one may check that in those cases there are no poles in the Cauchy integrals
over the moduli. 
The particle with charge $r,\tilde r$ corresponds to a scalar field $\varphi^{i,x}$  of mass $m_{r\tilde r}$ 
that would couple to the bi-adjoint $\Phi^{a \alpha}$ with a potential of the form
  \be
{\cal L }_{\text{int}} = -\frac{y_{r \tilde r}}{2} (T_r^a)^{ij} (\tilde T_{\tilde r}^\alpha)^{xy}
\Phi^{a \alpha} \varphi^{ix} \varphi^{jy} \;,
\ee
where by $T_r^a$ we indicate the generators in the symmetric tensor product of $r$ adjoint representations, and so on, compare with eq. \eqref{lag-phi3}. 
Note that such a representation has an index $T(r)=T(A)   {N_A+r \choose N_A+1} $,
which indeed is what appears in \eqref{76}. To make our formulas more explicit for the cases of
$SU(N)$ groups,  we  recall that $T(A)=N$ and $N_A = N^2-1$ in those cases. 

The methods we have developed in this section can be generalized to higher point correlation functions at 
1-loop, taking \eqref{pi-4} as starting point. The factorized structure of the two point function \eqref{self-ernegy-ex} is preserved at higher points so it is easy to see that our prescription \eqref{prescription-1} is enough 
to reproduce the $N$-point correlation functions contributing to the 1-loop Yang-Mills effective action starting from the $N$-point functions for the bi-adjoint particle. In Appendix \ref{higher-points} we present a Bern-Kosower-type of formula for an arbitrary number of vertex insertions.

\section{Multi-colored scalars and coupling to gauge  fields}
\label{extensions}
The action of the bi-adjoint particle suggests extensions that might also be of some interest. We will again 
proceed with extensions using  first the formulation with a matrix valued action, and then the one with
auxiliary variables. The choice of which formalism to use is a matter of taste, as both methods are equivalent, as previously described. 

\subsection{Path ordering and multi-adjoint scalars}
\subsubsection{Example: beta function for the multi-adjoint particle}
As an  extension, one may consider the $n$-adjoint scalar particle, quantum of the $n$-adjoint scalar field $\Phi^A=\Phi^{a_1 a_2 \cdots a_n}$
where the multi-index $A$ contains an adjoint index $a_i$  for each factor of the symmetry group $G_1\times G_2\cdots \times G_n$.
A suitable interacting QFT Lagrangian is given by   
\be
{\cal L} = \frac 12 \Phi^{A} (-\partial^2 + m^2) \Phi^{A} + \frac{y}{3!} F^{ABC} \Phi^{A} \Phi^{B} \Phi^{C} \;,
\label{n-ad} 
\ee
where 
\be
F^{ABC} = f^{a_1b_1c_1} f^{a_2b_2c_2} \cdots f^{a_n b_n c_n} \;.
\ee
The theory is nontrivial for $n$ even, otherwise the potential vanishes as  the $F^{ABC}$ are then completely antisymmetric.
In a quantum-background split $\Phi\to \phi+ \Phi$, the quadratic part of the quantum field Lagrangian  takes the same form of \eqref{43},
but with a matrix-like background potential 
\be
\hat \Phi^{AB}= F^{ABC}  \Phi^{C} \;.
\ee
The formula for the one-loop logarithmic divergences \eqref{div2} still applies, now with 
 \ba
 {\rm tr}\,  \hat \Phi^2 &= \alpha 
 \Phi^A \Phi^A
 \\
 {\rm tr}\,  \hat \Phi^3 &= \frac{\alpha}{2^n}  F^{ABC}  \Phi^{A} \Phi^{B} \Phi^{C}\;, 
 \label{cal}
  \ea
  where 
  \be
  \alpha  = \prod_{i=1}^n T_i(A)
  \ee
   is the product of the indices of the adjoint representation for each factor $G_i$. 
  Here we have taken $n$ to be an even integer.  Note that $\alpha$ is always a positive number. 
  These divergences are renormalized by adding to  the Lagrangian \eqref{n-ad} the counterterms
   \be
   {\cal L}_{\text{ct}}= (Z_\Phi-1) \tfrac12 \Phi^{A}  (-\partial^2) \Phi^{A} 
   +
   (Z_y-1) \tfrac{y}{3!}  F^{ABC} \Phi^{A} \Phi^{B} \Phi^{C} 
      \ee
with 
\be 
Z_\Phi-1 = - \frac{\alpha}{6} 
\frac{y^2}{(4\pi)^3} \frac{1}{\epsilon} \;, \qquad 
Z_y-1 = -\frac{\alpha}{2^n}  \frac{y^2}{(4\pi)^3} \frac{1}{\epsilon} \;,
\ee  
producing the one-loop beta function 
\be
\beta(y) = \frac{\alpha}{4} \left ( 1 - \frac{1}{2^{n-2}} \right )  \frac{y^3}{(4\pi)^3} \;.
\ee
Evidently,  the beta function for $n>2$ is positive, while it vanishes for $n=2$,
and it is negative for $n=0$  (setting $\alpha=1$ in such a case), consistently 
with the asymptotic freedom of the latter case. These results are in agreement with those
of ref. \cite{Gracey:2020baa}.

\subsubsection{Example: symmetric multi-adjoint scalar}
As a final example, let us consider a different theory for the  $n$-adjoint scalar by taking 
only groups of the type $SU(N)$ with $N\geq 3$, which admit the completely symmetric invariant 
tensor  $d^{abc}$. Models for a single symmetric tensor were introduced long time ago to study critical  
phenomena in condensed matter physics, see, e.g., refs. \cite{deAlcantaraBonfim:1980pe, deAlcantaraBonfim:1981sy} and  
refs. \cite{Gracey:2015tta, Gracey:2020tkk, Borinsky:2021jdb} for more recent discussions. Thus, let us consider the 
replacement in \eqref{n-ad}
\be
F^{ABC} \quad 
\to \quad  D^{ABC}  
= d^{a_1b_1c_1} d^{a_2b_2c_2} \cdots d^{a_n b_n c_n} \;, 
\ee
which is consistent for all integers $n\in \mathbb{N}$, as  $D^{ABC}$  is now always totally symmetric.
The corresponding one-loop effective action is again represented as in   \eqref{pi-new1}, but now with
\be
 \hat \Phi^{AB}= D^{ABC}  \Phi^{C} 
\ee
which produces 
 \ba
 {\rm tr}\,  \hat \Phi^2 &= x_1 \delta^{AB} 
 \Phi^A \Phi^B\;,
 \\
 {\rm tr}\,  \hat \Phi^3 &= x_2 D^{ABC}  \Phi^{A} \Phi^{B} \Phi^{C}\;, 
 \label{cal}
  \ea
with coefficients $x_1$ and $x_2$ that depend on the chosen groups.  They are computed by 
  \be
  x_{1} = \prod_{i=1}^{n}\frac{N_{i}^{2}-4}{N_{i}} \;, 	 \hskip .4cm x_{2} = \prod_{i=1}^{n}\frac{N_{i}^{2}-12}{2N_{i}} \;,
  \ee
 with $N_{i}$ being the dimension of the fundamental representation of each copy of the $SU(N_{i})$ group,
 see \cite{Haber:2019sgz} whose normalization for the $d^{abc}$ tensor we adopt.
 From  the one-loop logarithmic divergences  of the effective action \eqref{div2} 
    we extract the counterterms  
    \be
   {\cal L}_{\text{ct}}= (Z_\Phi-1) \tfrac12 \Phi^{A}  (-\partial^2) \Phi^{A} 
   +
   (Z_y-1) \tfrac{y}{3!}  D^{ABC} \Phi^{A} \Phi^{B} \Phi^{C} 
      \ee
with 
\be 
Z_\Phi-1 = - \frac{x_1}{6} 
\frac{y^2}{(4\pi)^3} \frac{1}{\epsilon} \;, \qquad 
Z_y-1 = - x_2 \frac{y^2}{(4\pi)^3} \frac{1}{\epsilon} \;,
\ee  
producing the one-loop beta function 
\be
\beta(y) =\frac{y^3}{(4\pi)^3} \left ( \frac{x_1}{4} -x_2 \right ) 
\ee
that is
\be
\beta(y) =\frac{y^3}{(4\pi)^3} \Bigg\{
\frac{1}{4} \left(\prod_{i=1}^{n}\frac{N_{i}^{2}-4}{N_{i}}	\right)  
\left[	1-\frac{1}{2^{n-2}}  \prod_{i=1}^{n}\frac{N_{i}^{2}-12}{N_{i}^{2}-4  }	\right]
\Bigg \} \,.
\ee
This formula includes the well-known asymptotic freedom case for $n=0$ (with $x_1=x_2=1$),
while for $n=1$ the result is consistent with \cite{Gracey:2015tta} upon mapping different conventions, 
showing a positive beta function for $N\leq4$, and a negative beta function (and thus asymptotic freedom) for $N\geq 5$. 
Further, the bi-adjoint particle ($n=2$) has beta function 
\be 
\beta(y) =  \frac{y^3}{(4\pi)^3} \frac{2}{N_{1}N_{2}}\left(N_{1}^{2}+N_{2}^{2}-16\right) \;, 
\ee
which is non-vanishing and always positive, independently of the dimension of the $SU(N)$ groups 
($N_i\geq 3$, as otherwise the $d^{abc}$ symbols would vanish).
Quite generally, the one-loop beta function is positive for $n\geq 2$ in these models.

\subsection{Multi-adjoint scalar using auxiliary variables}
Let us consider again  the case of a particle symmetric under the product of an arbitrary number 
of simple Lie groups, $G_1\times  G_2\times  \cdots \times G_n$,
with the wave-function transforming in an arbitrary representation $R_i$ with 
generators $(T^{a_i}_{R_i})^{\alpha_i}{}_{\beta_i}$ for each factor $G_i$.
 For that purpose we introduce 
 a set of color variables  $c_i^{\alpha_i}(\tau)$ and complex conjugate 
$\bar c^i_{\alpha_i}(\tau)$, with $i=1,\dots, n$, and indices that match those of the chosen representation,  
 to construct the color charges (no sum over the index $i$)
 \be
 C^{a_i}_i(\tau) = \bar c^i_{\alpha_i}(\tau)(T^{a_i}_{R_i})^{\alpha_i}{}_{\beta_i} c_i^{\beta_i}(\tau)
 \label{charge}
 \ee
 and write down the action for the phase space coordinates $(x,p, c_i, \bar c_i)$
and worldline gauge variables $(e, a_i)$ as follows 
\be
S =\int_0^1 \! d\tau 
\Big [p_\mu \dot x^\mu - \frac{e}{2} (p^2+m^2 +U(\Phi(x),C_i)) 
+\sum_{i=1}^{n}  ( \ii \bar \cc^i_{\alpha_i} (\partial_\tau +\ii a_i) \cc_i^{\alpha_i}  + s_i a_i ) \Big ]  \;.
\label{multi-adjoint}
\ee 
The potential $U(\Phi(x),C_i) $ is taken to depend on suitable scalar background fields, collectively denoted by $\Phi$, 
and color charges  $C_i$ as given above. Of course, the potential is required to be invariant under the global symmetry group.
Our previous bi-adjoint particle \eqref{action1} is a special case of this more general model with $n=2$, color variables $c_i, \bar c_i$
transforming in the adjoint representation of the corresponding group, and scalar potential $U(\Phi(x),C_i) $ 
taken as in \eqref{ham}.

Introducing gauge fields to the theory is straightforward. Starting with the phase space 
action \eqref{multi-adjoint} we add gauge field interactions by the minimal substitution 
\begin{align}
 p_\mu \to p_\mu-  \sum_i g_i A^i_\mu
\end{align}
in the  constraint, with  the non-abelian gauge fields
$A^i_\mu$ now given by (no sum over $i$)
\begin{align}
A^i_\mu=A_\mu^{a_i,i} (x)  C^{a_i}_i = A_\mu^{a_i,i}(x)\,  
\bar c^i_{\alpha_i}(\tau)(T^{a_i}_{R_i})^{\alpha_i}{}_{\beta_i} c_i^{\beta_i}(\tau) \;,
\end{align}
which also uses the composite color charge \eqref{charge}.
In order to gauge only a subset of the symmetry groups one may simply set the remaining  gauge couplings to zero. 
For instance, considering again the bi-adjoint case given by the
action \eqref{multi-adjoint} with $n=2$ and generators in the adjoint,
the coupling to a single gauge field can be implemented by, say, 
setting $g_1\equiv g$ and $g_2=0$.  Eliminating the momenta through their equations of motion, 
using $(T_{\rm A}^a)^b{}_c= -if^{abc}  $, and Wick rotating, 
produces the  action 
\begin{align}
\!\! S[x,\cc, \bar \cc, d, \bar d,  e, a, \tilde a; \Phi, A_\mu]= 
  \int_0^1 \!\!\dd \tau & \Bigg[  \frac12 e^{-1}
\dot x ^2 - g f^{abc} \dot x_\mu A^{\mu a}(x) \bar \cc^b  \cc^c 
+\frac{e}{2} (m^2 -y C^a  \Phi^{a\alpha}(x) {\tilde C}^{\alpha})  \label{action-gauge} \nonumber \\
&+ \bar \cc_a (\partial_\tau +\ii a) \cc^a  +  \bar d_\alpha (\partial_\tau +\ii \tilde a) d^\alpha  
 -\ii s (a +   {\tilde a})
\Bigg]\;,
\end{align}
where we used again the notation of previous sections. 
It is easy to see that the action \eqref{action-gauge} leads to 
two types of vertices, namely the spin one vertex $V^{(1)}[k,\varepsilon,a; x, c, \bar c ]$ and the spin zero vertex 
$V^{(0)}[k,a, \alpha; x, c, \bar c, d, \bar d]$ 
which must be considered separately (here $a$ denotes an adjoint color index, not the worldline gauge field).
Using dimensional reduction, similar types of interactions and corresponding
amplitudes have been considered in the context of the CHY representation \cite{Cachazo:2014xea} and  the double copy for 
supersymmetric theories \cite{Chiodaroli:2014xia, Chiodaroli:2017ngp}.  There, quartic scalar potentials are also included. Starting from our general approach these  potentials may be easily added to the theory.
Also, from the double copy dictionary \cite{Bern:2019prr}, one expects that the double copy 
procedure $C\to K$ at the level of the  vertex operators will lead to the Yang-Mills-Einstein theory.
To introduce the gravitational couplings in the worldline action one should covariantize  it   
by replacing  $\eta_{\mu\nu} \to g_{\mu\nu} (x)$ and implement the path
integral for curved spaces  \cite{Bastianelli:2006rx, Bastianelli:2011cc}.

\subsubsection{Example:  self-energy for the multi-adjoint scalar}

One can now 
repeat the exercise of Section \ref{Self-energy}, and calculate the one-loop self-energy correction to the propagator of 
the $n$-adjoint scalar due to particles that sits in the $r_i$ representation (meaning  the symmetric tensor product of 
$r_i$ copies of the adjoint) of the group $G_i$. Denoting by $N_{A,i}$ the dimension of the group $G_i$, we can write the self-energy as
\be
\Pi_{r}^{BC}(p) = -\frac{y_r^{2}}{2 (4\pi)^{\frac{D}{2}}} \delta^{BC}   
(P^{2})^{\frac{D}{2}-2}\Gamma\left(2-\tfrac{D}{2}\right) 
\prod_{i=1}^{n}
T_{i}(A)
\binom{N_{A,i} + r_i}{N_{A,i}+1}  \;,
\ee
where by $r$ we indicate the vector with components $r_i$. The definition of $(P^2)^x$ given 
in eq.~\eqref{74} now depends on a new mass $m_r$.
The case of the pure $n$-adjoint field propagating in the loop is obtained by setting  $r_{i}=1$ for all $i$. 
Again, the group theory factor $\prod_{i=1}^{n} T_{i}(A) \binom{N_{A,i} + r_i}{N_{A,i}+1} $
can be rewritten as the product of the indices of the $r_i$ representations, namely
$\prod_{i=1}^{n} T_{i}(r_i)$.
Similarly one could proceed as before to 
compute the beta functions for the couplings $y$ and $y_r$, showing in this case a nontrivial mixing.

\section{Conclusions}
\label{Conclusions}
In this paper we have employed worldline methods to study  the bi-adjoint scalar particle. In order to apply worldline methods we derived a
suitable particle action and used it to represent the 1-loop effective action of the corresponding QFT. We have
employed  two different approaches to deal with color degrees of freedom. 
On the one hand the simplicity of the double copy approach suggested the introduction of auxiliary bosonic color variables in order to 
keep track of path ordering. These auxiliary variables can be used also in coupling to gauge fields.
On the other hand we have  considered as well a direct path ordering, which we have exemplified
by computing the beta function of the theory.
We have verified the vanishing of the beta function at 1-loop in the  bi-adjoint scalar theory,
 as  noted recently in \cite{Gracey:2020baa}. More generally, we have obtained the 1-loop beta function for  the theory of a scalar field 
 with an arbitrary number of adjoint indices and with an interaction 
potential dictated by the structure constants of the symmetry groups, 
 finding again agreement with \cite{Gracey:2020baa}. 
 
Extensions of the model in the worldline approach are straightforward, and follow from the basic structure of the bi-adjoint case. 
It would be interesting to study the case of gravity and
color together in order to explore more deeply the double copy from the  worldline perspective.

Using the auxiliary variables prescription, in Section \ref{sec2} we have proposed a double copy prescription to lift vertex operators 
of the bi-adjoint scalar to those of gauge fields and gravity. 
This prescription leads to the effective action induced by a scalar particle coupled to a non-abelian gauge
field. It is closer in spirit to the formal substitution rules in the classical double copy \cite{Monteiro:2014cda}. Indeed, one of our motivations towards a worldline approach to the bi-adjoint
is aimed at applications at the classical level where the worldline formalism, in the  context of Wong's equations, can be useful to describe the dynamics of classical 
scattering \cite{delaCruz:2020bbn}. In particular, it would be interesting to derive worldline Feynman rules for Yang-Mills theory  
as done in refs.\cite{Mogull:2020sak, Jakobsen:2021smu, Jakobsen:2021lvp} for gravity.

We believe that our formulation of the multi-adjoint theory can be useful to study the color-kinematics duality and 
the double copy from a worldline perspective, with some properties already  noticed in \cite{Bastianelli:2012bz}
 and investigated more recently in  \cite{Casali:2020knc, Ahmadiniaz:2021fey},
perhaps pointing   towards a potential first principle understanding of its origin. 
In addition, the multi-adjoint particle could be used as a toy model to study
finite temperature systems with  color \cite{Jalilian-Marian:1999uob, Mueller:2019gjj, delaCruz:2020cpc}. 
Another interesting direction worth pursuing  
would be to study the worldline meaning of a novel color-kinematics relation,
suggested in \cite{Magnea:2021fvy}, where the momentum of the particle appearing in the plane wave 
of the vertex operator is substituted by color matrices.

\addsec{Acknowledgements}
LDLC  acknowledges financial support from the {\em Open Physics Hub}
at the Physics and Astronomy Department in Bologna. 
Some of our figures were produced with the help of TikZ-Feynman~\cite{Ellis:2016jkw}.

\appendix

\section{Seeley-DeWitt coefficients}
\label{appA}
Let us briefly review the perturbative calculation of the Seeley-DeWitt coefficients in \eqref{coeffs-SdW},
starting from the representation of the effective action in \eqref{pi-new2}, which contains the time ordering.
At the perturbative order needed to obtain eq. \eqref{coeffs-SdW}, the effect of the time ordering is 
inconsequential, but it is important at higher orders.

Using the split \eqref{3-split}, we Taylor expand the interaction term $S_{\text{int}} $ in \eqref{3-int}
around $x_0$
\be
S_{\text{int}}  =  T y \int_0^1 \!  \dd\tau\,  \left ( 
\hat \Phi(x_0) + 
q^\mu(\tau)\partial_\mu \hat \Phi(x_0) +\cdots
  \right ) \;,
  \ee
 insert it into the exponential in \eqref{pert-exp}, and  expand the exponential.
Then, collecting all terms of the same power in $T$,
and taking into account that the $q$-propagator
with the vanishing Dirichlet boundary conditions 
 goes like $T$
   \be
 \la q^{\mu}(\tau)q^{\nu}(\sigma)\ra = -2T \delta^{\mu\nu} \left ( \frac12 |\tau-\sigma| -\frac12 (\tau+\sigma) +\tau\sigma
 \right ) \;,
 \ee
 we easily find from the constant $\hat \Phi(x_0)$ term
  \be
a_0 = 1 \;, \quad 
a_1  = - y \hat \Phi(x_0) 
\;, \quad 
a_2 = \frac{y^2}{2} \hat \Phi^2(x_0)    \;, \quad 
a_3 =  -\frac{y^2}{6}   \hat \Phi^3(x_0)  + \cdots 
\ee
where the dots denote three additional terms arising from using the $q$-propagator. 
They are as follows. A first term is 
 \ba
 &\left \la {\bf T} \ \frac12 
  \left (T y \int_0^1 \!  \dd\tau\,  q^\mu(\tau)\partial_\mu \hat \Phi(x_0)  \right )
    \left (T y \int_0^1 \!  \dd\sigma\,  q^\nu(\sigma) \partial_\nu \hat \Phi(x_0)  \right )
    \right \ra 
 \\
 &= \frac12 T^2 y^2 \partial_\mu \hat \Phi(x_0)  \partial_\nu \hat \Phi(x_0) \int_0^1 \!  \dd\tau
 \int_0^1 \!  \dd\sigma\,   \la q^\mu(\tau) q^\nu(\sigma) \ra
 \\
 &= \frac{1}{12} T^3 y^2 (\partial_\mu \hat \Phi(x_0))^2 
 \ea
 where the time ordering  $ {\bf T} $ is again inconsequential as the matrices at different times commute.
 A second term  reads
 \ba
 &\left \la {\bf T} \ \frac12 
  \left (T y  \hat \Phi(x_0)  \right )
    \left (T y \int_0^1 \!  \dd\tau\,  \frac12 
    q^\mu(\tau) q^\nu(\tau) \partial_\mu \partial_\nu \hat \Phi(x_0) \right )
\right \ra 
\\
&= \frac14 T^2 y^2  \hat \Phi(x_0) \int_0^1 \!  \dd\tau\,   
    \la q^\mu(\tau) q^\nu(\tau)\ra  \partial_\mu \partial_\nu \hat \Phi(x_0) 
    \\
    &= \frac{1}{12} T^3 y^2  \hat \Phi(x_0) \partial^2 \hat \Phi(x_0) \;.
 \ea
A third term is similar, but with the matrices $\hat \Phi(x_0)$ and $\partial^2 \hat \Phi(x_0)$ 
interchanged. These matrices commute under the trace, so that this last
term just  doubles the previous one.
This completes the derivation of $a_3$, which is equivalent to the one in 
\eqref{coeffs-SdW} by adding a total derivative, that anyway drops out from the effective action.

\section{Effective vertices}
\label{higher-points}

Let us now discuss briefly how to get a general formula for the  1PI higher-point correlation functions at one-loop in 
momentum space, by taking \eqref{pi-4} as starting point.  The generalization of the plane wave expansion of the scalar field \eqref{back-2-pw} is 
given by
\be 
\Phi^{a\alpha}(x_{0}+q) =   \sum_{\ell=1}^{N}\delta^{a a_{\ell}}  \delta^{\alpha \alpha_{\ell}} e^{\ii p_{\ell} \cdot x_{0}} e^{\ii p_{\ell} \cdot q(\tau)} \;,
\ee
which can be inserted in \eqref{pi-4}, with the constraint that each plane wave should appear only once.
The integration over the loop base point produces a momentum conservation delta function which enables us to 
rewrite the 1-loop effective action as
\be
\Gamma[\Phi] = (2\pi)^{D}\delta^{D}\left( \sum_{\ell=1}^{N} p_{\ell}\right) \Gamma_{r \tilde{r}}^{a_{1} \alpha_{1} \cdots a_{N} \alpha_{N}}(p_{1}\cdots p_{n}) \;.
\ee 
Stripping off the momentum conservation Dirac delta function we define the 1PI $N$-point function as
\ba
&\Gamma_{r \tilde{r}}^{a_{1} \alpha_{1} \hdots a_{N} \alpha_{N}}(p_{1}\hdots p_{n}) = -\frac{1}{2}y_{r \tilde{r} }^{N}
\varointctrclockwise_{\gamma_{-}}\frac{dz}{2\pi \ii} \frac{1}{z^{r+1}(1-z)^{N_{A}}} 
\varointctrclockwise_{\gamma_{-}}\frac{dw}{2\pi \ii} \frac{1}{w^{\tilde{r}+1}(1-w)^{\tilde{N}_{A} }}\\
&\int_{0}^{\infty} \frac{dT}{T} \frac{e^{-m_{r \tilde{r}}^{2}T}}{ \left(4\pi T\right)^{\frac{D}{2}}} T^{N} \left( 
\prod_{\ell=1}^{N}\int_{0}^{1}d\tau_{\ell}  \right)
\Big\la e^{\ii \sum_{k=1}^{N} p_{k} q(\tau_{k})} \Big\ra  \, \Big\la \prod_{k=1}^{N} C^{a_{k}}(\tau_{k}) \Big\ra  
\Big \la \prod_{\ell=1}^{N} \tilde{C}^{\alpha_{\ell}}(\tau_{\ell}) \Big\ra \;,
\ea
where we also switched to the Wilson loop variables defined in \eqref{65}. 
The v.e.v of the kinematical part of the vertex operator can be evaluated 
using the identity 
\ba
\Big\la e^{\ii\sum_{k=1}^{N}p_{k}\cdot q(\tau_{k})}\Big\ra = e^{T\sum_{k,\ell=1}^{N}p_{k}\cdot p_{\ell}\del(\tau_{k},\tau_{\ell})}.
\ea
The integral over proper time $T$ can the be performed exactly 
\begin{align}
\int_{0}^{\infty}\!\! dT \, &e^{ -T \left(m_{r \tilde{r}}^{2} - \sum_{k,\ell=1}^{N} p_{k}\cdot p_{\ell} \del(\tau_{k},\tau_{\ell})		\right) }  T^{N-\frac{D}{2}-1}  \\
&= \Gamma\left(N-\tfrac{D}{2}\right)
\Bigg(m_{r \tilde{r}}^{2} - \sum_{k,\ell=1}^{N} p_{k}\cdot p_{\ell} \del(\tau_{k},\tau_{\ell})		\Bigg)^{\frac{D}{2}-N}. \nonumber
\end{align}
Finally, putting all of the pieces together we get our desired master formula
\ba
&\Gamma_{r \tilde{r}}^{a_{1} \alpha_{1} \cdots a_{N} \alpha_{N}}(p_{1}, p_{2},\hdots,p_{N}) =
-\frac{1}{2}\frac{y_{r \tilde{r}}^{N}}{(4\pi)^{\frac{D}{2}}}
 \varointctrclockwise_{\gamma_{-}} \frac{\dd z}{2\pi \ii} \, \frac{1}{z^{r+1}(1-z)^{N_{A}}}
 \varointctrclockwise_{\gamma_{-}}  \frac{\dd w}{2\pi \ii} \frac{1}{w^{\tilde{r}+1}(1-w)^{ \tilde{N}_{A}}  }\\
&\Gamma(N-\tfrac{D}{2})
\left(\prod_{\ell=1}^{N}\int_{0}^{1}d\tau_{\ell} \right)
\Bigg(
m_{r \tilde{r}}^{2}-\sum_{k,\ell=1}^{N}p_{k}\cdot p_{\ell}\, \del(\tau_{k},\tau_{\ell})
\Bigg) ^{\frac{D}{2}-N}
\Big\la \prod_{k=1}^{N} C^{a_{k}}(\tau_{k}) \Big\ra  
\Big \la \prod_{\ell=1}^{N} \tilde{C}^{\alpha_{\ell}}(\tau_{\ell}) \Big\ra ,
\ea
which holds in the case where the particle propagating in the loop has charge $r, \tilde{r}$ with respect to both Lie groups. The (amputated) external lines correspond instead to the plane waves of the bi-adjoint field $\Phi^{a\alpha}$. 

\bibliographystyle{JHEP}

 \renewcommand\bibname{References} 
\ifdefined\phantomsection		
  \phantomsection  
\else
\fi
\addcontentsline{toc}{section}{References}
\providecommand{\href}[2]{#2}\begingroup\raggedright\endgroup

\end{document}